\documentclass[preprint,showpacs,preprintnumbers,amsmath,amssymb]{revtex4}
\usepackage{graphicx}% Include figure files
\usepackage{dcolumn}% Align table columns on decimal point
\usepackage{bm}% bold math
\usepackage{epsfig}

\begin{document}

%\preprint{To appear in Physical Review A}
%PSU/TH-250 \\

\title{Self-interference of a single Bose-Einstein condensate
\\ due to boundary effects}

\author{R. W. Robinett} \email{rick@phys.psu.edu}
\affiliation{%
Department of Physics\\
The Pennsylvania State University\\
University Park, PA 16802 USA \\
}%

\date{\today}

\begin{abstract}

A simple model wavefunction, consisting of a linear combination of two 
free-particle Gaussians,  describes many of the observed features seen 
in the interactions of two isolated Bose-Einstein condensates as they 
expand, overlap, and interfere. We show that a simple extension of this 
idea can be used to predict the qualitative time-development of a single 
expanding BEC condensate produced near an infinite wall boundary, giving
similar interference phenomena. We also briefly discuss other possible 
time-dependent behaviors of single BEC condensates in restricted geometries,
such as wave packet revivals.

\end{abstract}

\pacs{03.75.Kk, 03.75.Hh, 03.75.Nt, 03.65.Ge}

\maketitle

\section{Introduction}
\label{introduction}

It can be argued that much of the early success of quantum theory can 
be traced to the fact that many exactly soluble quantum models are 
surprisingly coincident with naturally occurring physical systems, such 
as the hydrogen atom and the rotational/vibrational states of molecules.
Many other exemplary quantum mechanical models, which have historically
been considered as only textbook idealizations, have also 
recently found experimental realizations. Advances in areas such as materials 
science or laser trapping and cooling of atoms have allowed the production 
of approximations to a number of systems which had typically
been relegated to lists of pedagogical examples. While some such examples 
have found use as devices, many others have been applied to the study of 
fundamental quantum behavior.

For example, one-dimensional quantum wells (infinite and finite) are a 
staple of textbooks and have found use in modeling quantum dots and other
structures \cite{quantum_dots}, including the use of asymmetric wells
\cite{asymmetric_wells}. Two-dimensional quantum mechanical
`standing waves' have been observed in a wide variety of geometries
\cite{corrals}, while evidence for bound quantum states of 
the neutron in the Earth's gravitational field \cite{gravity}
(a problem which is often described as the `quantum bouncer' in the 
pedagogical literature) has recently been presented. The {\it ``generation 
of nonclassical motional states''}, such as coherent and squeezed states 
\cite{oscillator_states} of harmonically trapped ions,  has also been 
demonstrated and studied in detail as an example of time-dependent
designer wave functions in the most familiar of all model potentials.

The experimental realization of Bose-Einstein condensates (BECs) 
\cite{review_bec} has  allowed for an even wider variety of fundamental 
tests, including the 
``{\it Observation of interference between two Bose condensates}'' 
\cite{original_bec}. In the original experiment \cite{original_bec}, 
two samples of sodium atoms were evaporatively cooled
``{\it well below the transition temperature to obtain condensates}''
such that they were initially well-separated in a double-well potential. 
The two
condensates were then allowed to freely expand and interference effects were
observed in the resulting overlap region, while no similar effects were
seen for a single expanding condensate. Similar effects have been seen 
in other experimental realizations \cite{other_bec} and more recently have 
been observed with up to 30 uncorrelated condensates \cite{bec_many} 
produced in an optical trap.

As we will briefly review in Sec.~\ref{sec:simple_model}, a simple model
wavefunction consisting of a linear combination of two \cite{wallis} 
(or more \cite{bec_many}) Gaussian terms, $\psi_{(G)}(x,t;x_0$), 
(one for each condensate) captures many of the salient features observed 
experimentally. A (single particle) wavefunction of the form
\begin{equation}
\psi(x,t) = N \left[\psi_{(G)}(x,t;x_A) + e^{i\phi} \psi_{(G)}(x,t;x_B)
\right]
\end{equation}
can be used, with $x_0=x_A,x_B$ describing the locations of the two isolated
condensates and with fixed relative phase \cite{anderson} (and a 
normalization constant $N$) and we will
discuss the predictions of this simple model in the next section.

Such linear combination solutions (especially of Gaussians) have been
frequently used in the pedagogical literature to describe the 
time-development of wave packet solutions of the 1D Schr\"{o}dinger 
equation (SE) describing an otherwise free-particle impinging on an infinite 
wall or barrier (or ``free particle on the half-line''.) 
``Mirror'' or ``image'' solutions of the form
\begin{equation}
      \tilde{\psi}(x,t) = \left\{ \begin{array}{ll}
       \tilde{N} \left[\psi(x,t)-\psi(-x,t)\right] & \mbox{for $x\leq 0$} \\
               0 & \mbox{for $0\leq x$}
                                \end{array}
\right.
\label{image_mirror_solution}
\end{equation}
(for a particle restricted to $x\leq 0$) satisfy the 1D 
free-particle SE (if $\psi(x,t)$ does) and also automatically satisfy
the appropriate boundary condition at the infinite wall (assumed to
be at $x=0$) for all times.  
Such solutions have been used in a variety of pedagogical 
applications \cite{andrews} -- \cite{thaller}, but also 
in a research context to discuss the deflection of ultracold quantum 
particles (wavepackets) from impenetrable boundaries or mirrors 
\cite{dodonov}. Such analyses naturally explain the spatially oscillatory
behavior of the position-space probability density as the wave packet 
`hits' the wall (as observed in numerical calculations) as the interference 
between two overlapping terms, much like the observed BEC effect.

In this Letter, we note that a single BEC condensate, produced
near an infinite wall boundary and allowed to expand freely, will likely
exhibit interference effects describable by such 
`mirror' or `image' solutions as in Eqn.~(\ref{image_mirror_solution}), and we
discuss this in Sec.~\ref{sec:extensions}. We also extend such
ideas to a localized BEC produced between two infinite boundaries (as in an 
infinite well potential) to very briefly discuss other possible effects, 
such as wave packet revivals.

\section{Simple model of interfering Bose-Einstein condensates}
\label{sec:simple_model}

A description of the initial (single particle) 
one-dimensional wavefunction for two
separated BEC condensates using isolated Gaussian forms has been 
made by the authors of Ref.~\cite{wallis}, who then analyze the 
resulting time-development
of the BECs using a Wigner distribution approach. (A generalization
to multiple BECs is given in Ref.~\cite{bec_many}.)
As a review of such an approach and its successes in qualitatively modeling 
more sophisticated analyses (and the experimental data), consider two BEC
condensates, separated by a distance $d$ and initially centered at 
$x_0 = \pm d/2$, described by the time-dependent free-particle wavefunction
\begin{equation}
\psi_{2BEC}(x,t) = \frac{N}{\sqrt{\sqrt{\pi} \beta (1+it/t_0)}}
\left[ 
e^{-(x-d/2)^2/2\beta^2(1+it/t_0)}
+
e^{i\phi}
e^{-(x+d/2)^2/2\beta^2(1+it/t_0)}
\right]
\label{two_gaussian_solution}
\end{equation}
where the normalization factor is given by
\begin{equation}
N = \frac{1}{\sqrt{2}}\left(1 + \cos(\phi) e^{-d^2/4\beta^2}\right)^{-1/2}
\, . 
\end{equation}
The time-dependent spatial width for a {\it single} such Gaussian term 
is given by
\begin{equation}
\Delta x_t = \frac{\beta_t}{\sqrt{2}}
\equiv  \frac{\beta}{\sqrt{2}}\sqrt{1+(t/t_0)^2}
\qquad
\quad
\mbox{where}
\quad
\qquad
t_0 \equiv  \frac{m\beta^2}{\hbar}
\end{equation}
and the corresponding spread in momentum-space (again, for a single Gaussian)
is $\Delta p_t = \Delta p_0 = \hbar/(\beta\sqrt{2})$. The time-dependent
probability density for the two condensate state is then given by
\begin{equation}
P_{2BEC}(x,t) = \frac{N^2}{\sqrt{\pi} \beta_t}
\left[
e^{-(x-d/2)^2/\beta_t^2}
+
e^{-(x+d/2)^2/\beta_t^2}
+ 2 e^{-(d^2+4x^2)/4\beta_t^2}
\cos\left(\phi + \frac{tdx}{t_0 \beta_t^2}\right)
\right]
\end{equation}
where the cross-term describes the interference effect.

For each Gaussian contribution, there are momentum components of order
$p \sim \Delta p_0 = \hbar/\beta\sqrt{2}$  so that the time ($T_{O}$) 
it takes such components to drift from one condensate and overlap with 
the other is of order $T_{O} \sim d/(p/m) \sim \sqrt{2} dm\beta/\hbar$.
For condensates which are initially highly localized and well-separated,
the time dependent position-spread is then dominated by the $(t/t_0)^2$
term since $ (T_{O}/t_0)^2 \sim 2(d/\beta)^2 >> 1$. In that limit, 
the oscillatory term is then approximately given by
\begin{equation}
\cos\left(\phi + \frac{xdt}{\beta^2_t t_0}\right)
\qquad
\longrightarrow
\qquad
\cos\left(\phi + \frac{xdt_0}{\beta^2 t}\right)
\end{equation}
so that the local wavelength variations seen in the interference pattern
are time-dependent and scale like
\begin{equation}
\frac{2\pi}{\lambda}x  = kx = \frac{xdt_0}{\beta^2t}
\qquad
\qquad
\mbox{or}
\qquad
\qquad
\lambda = \frac{h t}{md}
\, , 
\end{equation}
just as in Eqn.~(1) of Ref.~\cite{original_bec}. The time-dependent real and 
imaginary parts of the individual components of this simple wavefunction 
are nicely consistent with more detailed calculations \cite{other_bec}
where the BEC is ``{\it characterized by a phase that varies quadratically
...across the condensate}''; for example, compare the pedagogical 
illustration of an expanding  $p=0$ Gaussian wavepacket in Fig.~2
of Ref.~\cite{foundations_paper} with Fig.~1 of Ref.~\cite{other_bec}. 
Finally, the corresponding momentum-space probability density is given by
\begin{equation}
|\phi_{2BEC}(p,t)|^2
= \frac{4N^2\alpha}{\sqrt{\pi}}
\cos^2\left(\frac{pd}{2\hbar}\right)
e^{-\alpha^2p^2}
\end{equation}
where $\alpha \equiv \beta/\hbar$ so that there is indeed structure
in momentum space at integral multiples of $p = h/d$, as discussed in
more detail in Ref.~\cite{bec_momentum}. Thus, in many important ways,
the simple wavefunction in Eqn.~(\ref{two_gaussian_solution}) encodes 
much of the physics observed in the interference of two expanding
BEC condensates.

\section{Single BEC near a infinite wall boundary 
and related effects}
\label{sec:extensions}

Motivated then by earlier pedagogical papers on `mirror' or
`image' solutions \cite{andrews} -- \cite{thaller}, 
we can imagine a single BEC condensate produced close to an infinite
barrier. If the barrier is located at $x=0$ and the single condensate 
is produced at $x=-d/2$, the resulting `mirror' solution 
in Eqn.~(\ref{image_mirror_solution}), for $x \leq 0$, 
can be described by the form in Eqn.~(\ref{two_gaussian_solution}) 
with $\phi = \pi$ (so that $\cos(\phi) = -1$) and with normalizations
simply related by $\tilde{N} = \sqrt{2}N$. The resulting time-dependent
solution will then exhibit the same type of interference patterns observed
for two isolated condensates. One possibility for such an infinite barrier 
might be an atomic mirror \cite{mirror_ideas} of the type 
successfully used in a number of atomic physics applications 
\cite{mirror_applications}.

The addition of a second infinite wall barrier (say at $x  = -d$)
to such a case might then be modeled by the standard infinite well
problem of textbooks. The time development of wave packet propagation
in this system can then be described in terms of an infinite number
of image solutions \cite{infinite_well} 
and discussions related to this approach go back to at least Einstein 
and Born \cite{born}. For a single BEC condensate in this restricted
geometry, modeled as a $p=0$ Gaussian, 
in addition to the spreading/coherence time ($t_0$) and the time
to overlap the other real or image condensate ($T_{O}$), the only other
relevant time scale is the quantum revival time \cite{robinett_revival}
$T_{rev}$. 
For the quantized energy eigenvalues in an infinite well of width $d$ 
(as imagined here) the revival
time for an arbitrary localized wave packet is $T_{rev} = 4md^2/\hbar \pi$
\cite{segre}.
The ratio of revival time to overlap time is then $T_{rev}/T_{O}
\sim 2d/\pi\Delta x_0 >>1$. In the original two BEC experiment 
\cite{original_bec}, the two condensates are allowed to fall freely
as they expand, so a more detailed analysis of a specific experiment 
realization would be required to determine if the revival time is
too long to be observed. One should note, however, that for the special
case of a $p=0$ Gaussian waveform produced precisely in the center of such 
an infinite 
well, because only even energy eigenstates are excited, the effective revival
time is actually $T_{rev}/8$ \cite{robinett_revival} 
for this very special geometry, which gives almost an order-of-magnitude 
shorter fall time to achieve a revival.

Other arrangements of two infinite plane barriers can also be imagined to
give rise to interesting BEC interference effects, which can be modeled
using `mirror' or `image' methods. For example, two such infinite walls
can be placed at right angles, to form a `corner ($90^{\circ}$) reflector', 
defined by the potential
 \begin{equation}
V(x,y) = \left\{ \begin{array}{cl}
               0 & \mbox{for $0<x$ {\bf  \underline{and}} $0<y$} \\
               +\infty  & \mbox{otherwise}
                                \end{array}
\right.  
\, . 
\end{equation}
A solution of the form
\begin{equation}
\psi_{corner}(x,y;t) = N\left[ \psi(x,y;t) - \psi(-x,y;t) - \psi(x,-y;t)
+ \psi(-x,-y;t)\right]
\end{equation}
making use of three auxiliary `image' components, solves the Schr\"{o}dinger
equation in the allowed region for any free-particle solution $\psi(x,y;t)$, 
as well as  satisfying  the boundary
conditions at the two walls. The normalization factor can be obtained
explicitly in the case of an Gaussian solution. as above. 
The same construction can also be employed for other angles between the 
two walls, $\Theta$, using familiar examples from optics, such as for the
cases of $\Theta = 45^{\circ}$ and $60^{\circ}$.

\end{document}